\documentclass[twocolumn,showpacs,amsmath, preprintnumbers,amssymb, aps, prl]{revtex4-1}
\usepackage{graphicx}
\usepackage{dcolumn}
\usepackage{bm}

\begin{document}

\preprint{deposit manuscript by VINITI, 03 May 2012, N205-Â2012(in Russia)}

\title{Casimir expulsion of periodic configurations}

\author{Evgeny\,G.\,Fateev}
 \email[Electronic address: ]{e.g.fateev@gmail.com}
\affiliation{%
Institute of mechanics, Ural Branch of the RAS, Izhevsk 426067, Russia
}%
\date{\today}

\begin{abstract}
There is the possibility in principle that the noncompensated Casimir force 
exists in open nanosized metal cavities arranged in the form of periodic 
structures. It is found that when trapezoid cavities are strictly periodic 
all the Casimir expulsion forces are completely compensated. However, when 
the distance of the gap is formed between the cavities, in the periodic 
configuration a noncompensated expulsion force proportional to the number of 
cavities appears. There are such effective parameters of the periodic 
configuration (the angles of the opening of cavities, their lengths and the 
relationships between them) which lead to the appearance of a maximum of 
expulsion forces per unit of structure length.
\end{abstract}

\maketitle

In Ref.~\cite{Fateev:2012} the possibility in principle is shown that the 
noncompensated Casimir force can exist in open nanosized metal cavities. The 
effect is theoretically demonstrated for a single trapezoid configuration. 
The force manifests itself as the time-constant expulsion of open cavities 
in the direction of their least opening. The optimal parameters of the 
angles of opening (broadening) of the cavities' generetrices and their 
lengths are found, at which the expulsion force is maximal. It should be 
noted that the force differs significantly from expulsion forces capable of 
creating effects of levitation-type over bodies-partners~\cite{Jaffe:2005,
 Leonhardt:2007, Levin:2010, Rahi:2010, Rahi:2011}. 
The question arises if the existence 
of noncompensated expulsion forces is possible in periodic structures based 
on trapezoid configurations possessing the effect of expulsion. A particular 
case of trapezoid configurations is Casimir parallel mirrors 
\cite{Casimir:1948, Casimir:1949} which do not possess an 
effective expulsion force \cite{Fateev:2012}. 

Let us consider a periodic configuration with trapezoid cavities as an 
illustration of the possibility of the Casimir expulsion force existence. 
Note that a single cavity is understood as an open thin-walled metal shell 
with one or several outlets. The inner and outer surfaces of the cavity 
should have the properties of perfect mirrors. The cavity should entirely be 
immerged into a material medium or be a part of the medium with the 
parameters of dielectric permeability being different from those of physical 
vacuum. In Cartesian coordinates the configuration looks like two thin metal 
plates with the surface width $L$ (oriented along the $z$-axis) and length R, 
which are situated at a distance $a$ from one another; the angle $2\varphi $ of 
the opening of the generating lines of cavities between the plates can be 
varied (by the same value $\varphi $ imultaneously for both wings of the 
trapezoid cavity) as it is shown in Fig.\hyperlink{fig1}{1}. 
\begin{figure}[htbp]
\hypertarget{fig1}
\centerline{\includegraphics[width=1.4in,height=1.6in]{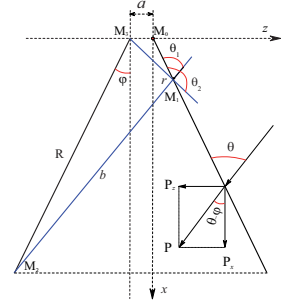}}
 \caption{Schematic view of the configuration of a symmetric trapezoid cavity 
with the length $R$ of the wing surface, a particular case of which is parallel 
planes, i.e. $\varphi =0$, and a triangle at $a=0$. The section of the cavity 
shown in the Cartesian coordinates in the plane $(x,z)$ has the width $L$ in the 
$y$ direction normal to the plane of the figure. The blue straight lines 
designate virtual rays with the length $b$ coming from the point $\mbox{M}_1 $at 
limit angles $\Theta _1 $ and $\Theta _2 $ onto the right cavity surface 
ending at the ends of the opposite cavity wing at the points $\mbox{M}_2 $ 
and $\mbox{M}_3 $, respectively. }
\end{figure}
\begin{figure}[htbp]
\hypertarget{fig2}
\centerline{\includegraphics[width=2.5in,height=1.0in]{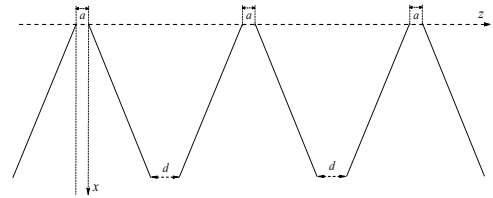}}
 \caption{Schematic view of the periodic 
configuration, in which the broadened parts of the symmetrical trapezoid 
cavities with the distance of the gap $d$ between them are in the x-direction.}
\end{figure}

The periodic configuration with trapezoid figures in the Cartesian 
coordinates looks like it is shown in Fig. \hyperlink{fig2}{2}. Each figure in the 
configuration period is similar to a single figure in Fig. \hyperlink{fig1}{1}. Between the 
ends of the figures in the period there is a distance of the gap $d$.

For each figure the expulsion force in the $x$-direction can be found in the 
first approximation in the form \cite{Fateev:2012} 
\begin{equation}
\label{eq1}
F_x =\int\limits_0^L {dy} \int\limits_0^R {P_x (\varphi ,\Theta ,r)} \,dr.
\end{equation}
Here, the local specific force of expulsion at each point $r$ on the cavity 
wing with the length $R$ and width $L$ is
\begin{equation}
\label{eq2}
\begin{gathered}
  {P_x}(r) = \frac{{\hbar c{\pi ^2}}}{{{{240}^{}}{s^4}}}\int\limits_{{\Theta _1}}^{{\Theta _2}} {} \sin {(\Theta  - 2\varphi )^4}\cos (\Theta  - \varphi )d\Theta  \\ 
   =  - \frac{{\hbar c{\pi ^2}}}{{{{240}^{}}{s^4}}}{A}(\varphi ,{\Theta _1},{\Theta _2}), \\ 
\end{gathered}
\end{equation}
where
\begin{equation}
\label{eq3}
\begin{gathered}
A (\varphi ,\Theta _1 ,\Theta _2 ) \\ 
 =\frac{1}{240}\Bigl[ 90\sin (\varphi -\Theta _1 )-90\sin (\varphi -\Theta _2 ) \\ 
 +60\sin (3\varphi -\Theta _2 )-60\sin (3\varphi -\Theta _1 ) \\ 
 +20\sin (5\varphi -3\Theta _2 )-20\sin (5\varphi -3\Theta _1 ) \\ 
 +5\sin (7\varphi -3\Theta _1 )-5\sin (7\varphi -3\Theta _2 ) \\ 
 +3\sin (9\varphi -5\Theta _1 )-3\sin (9\varphi -5\Theta _2 )\Bigr].
\end{gathered}
\end{equation}
In formula (\ref{eq2}), $\hbar =h/2\pi$  is reduced Planck constant, c is light 
velocity, and the functional expressions for limit angles $\Theta _1$,
$\Theta _2$ (see Fig. \hyperlink{fig1}{1}) in the trapezoid cavity and the parameter $s$ are
\begin{equation}
\label{eq4}
\begin{gathered}
\Theta_1 =\mbox{arccos}\Biggl\{-(r+a\sin \varphi - R\cos 2\varphi )\\
\times\Bigl[(a+R\sin \varphi +r\sin \varphi)^2\\
 +(r\cos \varphi -R\cos \varphi )^2 \Bigr]^{-\frac{1}{2}}\Biggr\},
\end{gathered}
\end{equation}
\begin{equation}
\label{eq5}
\Theta _2 =\mbox{arccos}\left[ {-\frac{r+a\sin \varphi }{\sqrt 
{a^2+r^2+2ra\sin \varphi } }} \right],
\end{equation}
and 
\begin{equation}
\label{eq6}
s=\frac{\sin (2\varphi -\Theta _2 )(a+r\sin \varphi )}{\sin (\varphi -\Theta _2)}.
\end{equation}

When such cavities are being arranged in the periodic structure, the 
following should be kept in mind. The periodic arrangement of $n$ 
trapezoid cavities being at the distance $d$ from one another, which sides with 
the widest opening are directed against the $x$ axis, leads to the formation of 
$n$-1 cavities with oppositely directed openings (see Fig. \hyperlink{fig2}{2}). In this case, a 
wing (one of the surfaces of the trapezoid cavity) of each cavity is a wing 
of the other cavity, the opening of which is oppositely directed. Thus, for 
$n$ cavities periodically arranged along the $y$ axis we can write the expression 
for the total force of expulsion along the$ x$ axis
\begin{equation}
\label{eq7}
F_{\sum} =nF_x (a)-(n-1)F_x (d).
\end{equation}
Here, $F_x (a)$ is the force along the $x$ axis for the distance $a$ between the 
nearest ends of the cavities, and $F_x (d)$ is, respectively, the force for 
the distance of the gap $d$ between periods instead of $a$ in formulas (\ref{eq1}-\ref{eq6}). From 
formula (\ref{eq7}) it is clear that for $d = a$ in the strictly periodic 
configurations the expulsion force is $F_{\sum} \to F_x (a)$ at $n\to \infty $.
That is even at $n\to \infty $ the thrust force in a strictly periodic 
structure always remains at the level of expulsion forces for a single 
cavity and is directed to the least opening of the cavity wings. It means 
that in the configuration the thrust force will be created which is directed 
against the $x$-direction. However, it is clear that at $d\ne a$ the expulsion 
force of the periodic configuration will not remain at the same level and 
will depend on the $d/a$ relation according to formula (\ref{eq7}) for different 
angles $\varphi $ of the opening of cavities as it is shown in Fig.\hyperlink{fig3}{3($a$)}. When the 
number $n$ of trapezoid cavities in the periodic configuration is growing, the 
character of the curves will be similar to that of the curves presented; 
however, of course, their level along the coordinate $F_x $ for any angles 
$\varphi $ and parameters will grow linearly.
\begin{figure*}
\hypertarget{fig3}
\centerline{
\includegraphics[width=1.6in,height=1.6in]{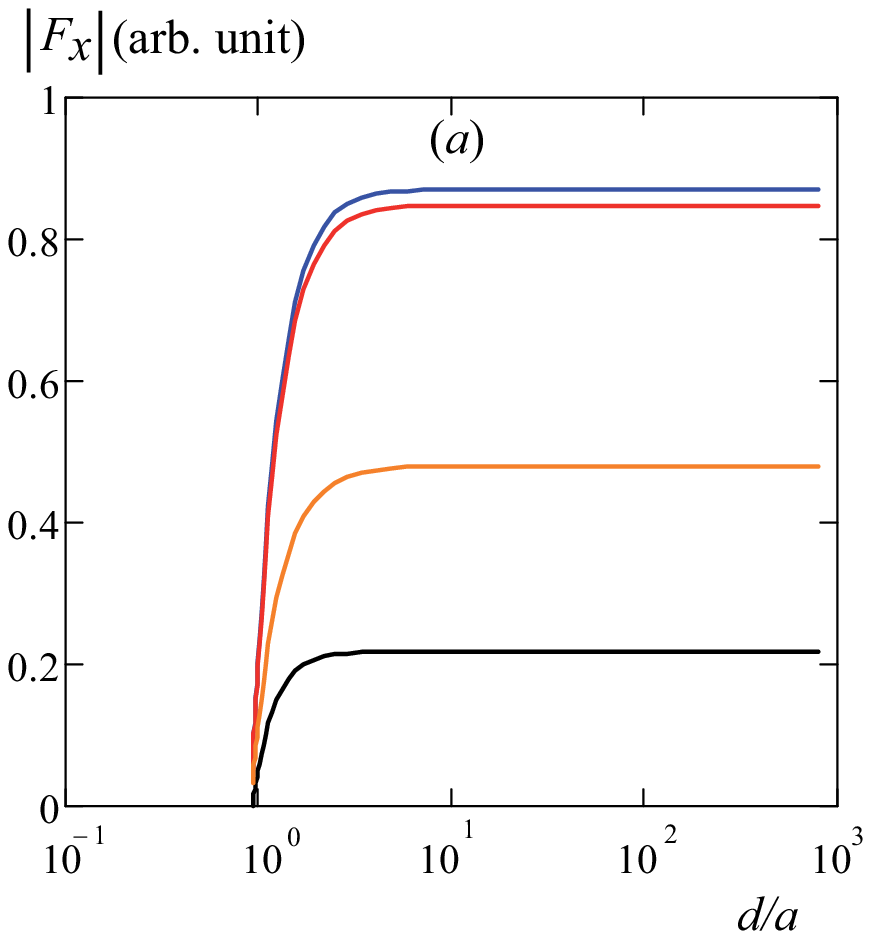}
\includegraphics[width=1.6in,height=1.6in]{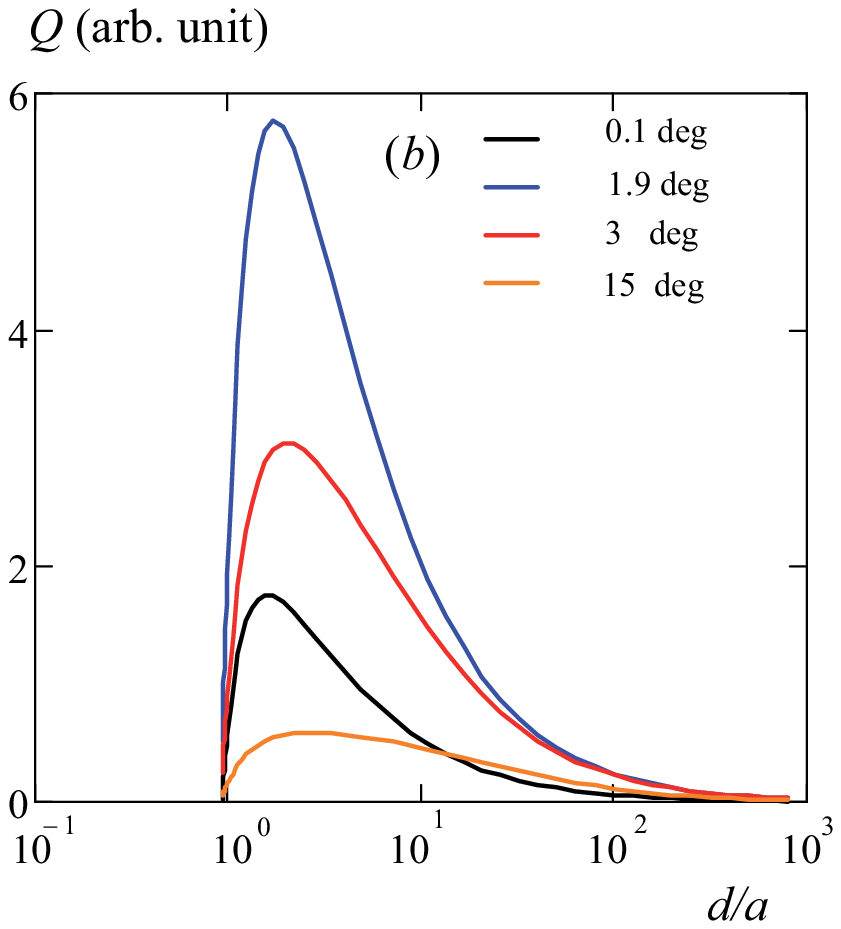}
\includegraphics[width=1.6in,height=1.6in]{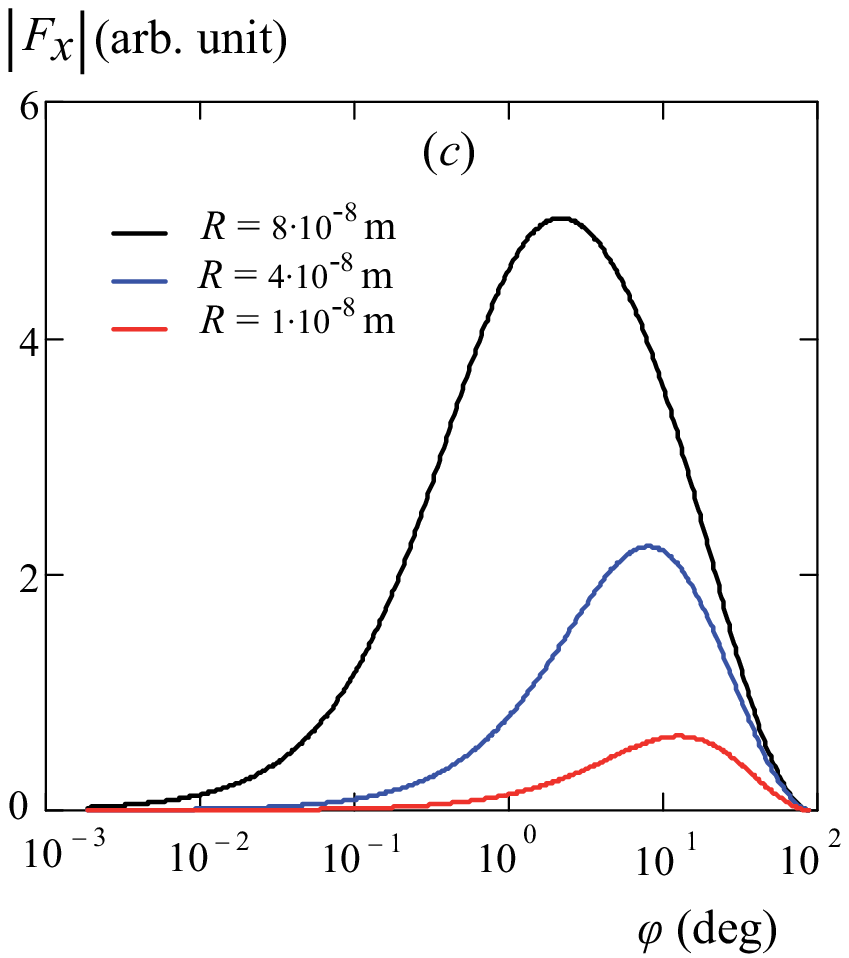}}
 \caption{Absolute values of the Casimir expulsion forces in the periodic 
configuration in the direction opposite to the $x$-axis depending on the 
relation $d/a$ in the structure ($a$) and the effectiveness $Q$ of expulsion of 
the structure ($b$) at different angles $\varphi $ of the opening of the cavities. 
The total Casimir force of expulsion ($c$) for different lengths $R$ depending on 
the angle $\varphi $.}
\end{figure*}

It is possible to determine the effectiveness $Q$ of the expulsion of $n$ 
cavities as the relation of the total force $F_{\sum}$ to the entire length 
of the configuration along the $y$ axis
\begin{equation}
\label{eq8}
Q=\frac{F_{\sum} }{n\left( {a+2R\tan \varphi} \right)+(n-1)d}.
\end{equation}
The dependence of $Q$ on the $d/a$ relation is displayed in Fig.\hyperlink{fig3}{3($b$)}. It can be 
seen, for example, that at $a=4\times 10^{-9}$\, m, for any length of the 
cavity wings $R$ with different angles $\varphi $, there is a maximum of 
effectiveness $Q$ of expulsion. As is known \cite{Fateev:2012} there 
is a maximum of the expulsion forces for each trapezoid figure depending on 
the angle of the opening of the cavities' wings [Fig. \hyperlink{fig3}{3($c$)}] and their lengths. 
In the periodic configurations with the distance of the gap $d$, there is a 
maximum of the expulsion effectiveness $Q$ as well. The maximum of 
effectiveness, which is common for two parameters $\varphi $ and $d/a$, is 
shown in Fig. \hyperlink{fig4}{4}. It was found that for $a=4\times 10^{-9}$\, m and $R/a=2.5$, 
the best angle is $\varphi \approx 5.59\;\mbox{deg}$ at $d/a\approx 1.58$ and 
$n$ = 2. For the given relation $R/a$ at $n\to \infty $ the angle is $\varphi \to 
6.0\;\mbox{deg}$ and $d/a\to 1.7$. At $R/a\to 20$ and $n\to \infty $, the 
angle $\varphi \to 0.1\;\mbox{deg}$ and $d/a\to 1.85$. 
\begin{figure}[htbp]
\hypertarget{fig4}
\centerline{\includegraphics[width=2.4in,height=1.8in]{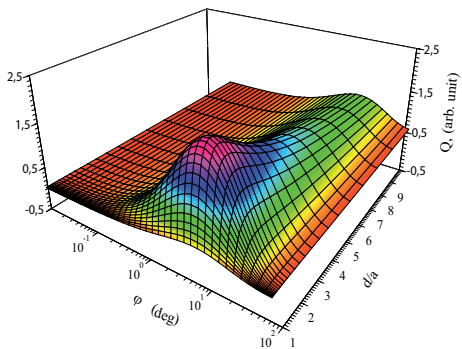}}
 \caption{Effectiveness $Q$ of the Casimir expulsion of the periodic structure 
depending on the angle $\varphi $ and relation $d/a$.}
\end{figure}

Note that when the trapezoid cavities are chequer-wisely arranged in the 
periodic configuration, i.e. the cavity with the widening of the opening 
against the $x$ axis is put to the cavity with the wide opening in the 
$x$-direction (so that the boundary wings become common to both cavities), in 
the system there will be a torque moment around the centre of mass of the 
configuration even at $d=a$. Of course, at certain combinations of the 
arrangement of cavities in the periodic configuration, a much larger torque 
can be achieved at $d\ne a$ compared to that at $d=a$.

Thus, in the present paper, the possibility in principle is shown that the 
noncompensated Casimir force exists in open nanosized metal cavities 
arranged in the form of periodic structures. It is found that in strictly 
periodic structures based on trapezoid figures all the Casimir expulsion 
forces are practically completely compensated. However, when the distance of 
the gap is formed between the cavities, in the periodic configuration a 
noncompensated expulsion force appears. In this case, at any relations of 
the configuration parameters (angles of opening and wing length of the 
cavities, the distance between the cavities, etc.) and at any number of 
cavities in the period there is an effective maximum of the expulsion 
forces. In some periodic structures there can be a torque moment around the 
centre of mass of the configuration.
\begin{acknowledgments}
The author is grateful to T. Bakitskaya for his helpful
participation in discussions.
\end{acknowledgments}

\end{document}